\newcommand{\beq}{\begin{equation}}
\newcommand{\eeq}{\end{equation}}
\newcommand{\bea}{\begin{eqnarray}}
\newcommand{\eea}{\end{eqnarray}}

\newcommand{\gsim}{\lower.7ex\hbox{$\;\stackrel{\textstyle>}{\sim}\;$}}
\newcommand{\lsim}{\lower.7ex\hbox{$\;\stackrel{\textstyle<}{\sim}\;$}}

\newcommand{\mrm}{\mathrm}

\documentclass[aps,prev,twocolumn,preprintnumbers,floatfix,nofootinbib]{revtex4-1}
\pdfoutput=1
\usepackage{geometry,amsmath,amsfonts}
\usepackage{graphicx}
\usepackage{epstopdf}
\usepackage{mathrsfs}
\usepackage{amssymb}
\usepackage{verbatim}
\usepackage{color}
\usepackage{multirow}

\def\stacksymbols #1#2#3#4{\def\theguybelow{#2}
    \def\vp{\lower#3pt}
    \def\sp{\baselineskip0pt\lineskip#4pt}
    \mathrel{\mathpalette\intermediary#1}}

\def\intermediary#1#2{\vp\vbox{\sp
     \everycr={}\tabskip0pt
     \halign{$\mathsurround0pt#1\hfil##\hfil$\crcr#2\crcr
              \theguybelow\crcr}}}

\def\be{\begin{equation}}
\def\ee{\end{equation}}
\def\bea{\begin{eqnarray}}
\def\eea{\end{eqnarray}}

\def\sp{\;\;\;,\;\;\;}

\def\mrm{\mathrm}

\def\lsim{\raise0.3ex\hbox{$\;<$\kern-0.75em\raise-1.1ex\hbox{$\sim\;$}}}
\def\gsim{\raise0.3ex\hbox{$\;>$\kern-0.75em\raise-1.1ex\hbox{$\sim\;$}}}

\def\inbar{\,\vrule height1.5ex width.4pt depth0pt}

\def\IC{\relax\hbox{$\inbar\kern-.3em{\rm C}$}}
\def\IQ{\relax\hbox{$\inbar\kern-.3em{\rm Q}$}}
\def\IR{\relax{\rm I\kern-.18em R}}
 \font\cmss=cmss10 \font\cmsss=cmss10 at 7pt
\def\IZ{\relax\ifmmode\mathchoice
 {\hbox{\cmss Z\kern-.4em Z}}{\hbox{\cmss Z\kern-.4em Z}}
 {\lower.9pt\hbox{\cmsss Z\kern-.4em Z}}
 {\lower1.2pt\hbox{\cmsss Z\kern-.4em Z}}\else{\cmss Z\kern-.4em Z}\fi}

\def\comment#1{}

\def\u1x{U(1)_X}
\newcommand{\nc}{\newcommand}
\nc{\LL}{L}
\nc{\vv}{\tilde{v}}
\nc{\ccdot}{\!\cdot\!}
\nc{\gsm}{G_{SM}}
\nc{\vfive}{\mathbf{5}\oplus\mathbf{\overline{5}}}
\nc{\vten}{\mathbf{10}\oplus\mathbf{\overline{10}}}
\nc{\zhol}{Z^{\rm hol}}
\nc{\xfb}{\,{\rm fb}}

\begin{document}

%
%

\preprint{LPT--Orsay 13-12}
\preprint{IFT-UAM/CSIC-13-014}
\preprint{UMN--TH--3137/13}
\preprint{FTPI--MINN--13/04}

\vspace*{1mm}

\title{Z-portal dark matter}

\author{Giorgio Arcadi$^{a,c}$}
\email{giorgio.arcadi@th.u-psud.fr}
\author{Yann Mambrini$^{a}$}
\email{yann.mambrini@th.u-psud.fr}
\author{Francois Richard$^{b}$}
\email{richard@lal.in2p3.fr}

\vspace{0.1cm}
\affiliation{
${}^a$ Laboratoire de Physique Th\'eorique 
Universit\'e Paris-Sud, F-91405 Orsay, France.}
\affiliation{
${}^b$ Laboratoire de l'Acc\'el\'erateur Lin\'eaire, IN2P3/CNRS and Universit\'e Paris-Sud 11
Centre Scientifique d'Orsay, B. P. 34, F-91898 Orsay Cedex, France.}
\affiliation{
${}^c$ 
 Institute for Theoretical Physics, Georg-August University G\"{o}ttingen, Friedrich-Hund-Platz 1,
 G\"{o}ttingen, D-37077 Germany
 }

\begin{abstract} 

\noindent
We propose to generalize the extensions of the Standard Model where the $Z$ boson serves as a mediator between the Standard
Model sector and the dark sector $\chi$. We show that, like in the Higgs portal case, the combined constraints from the 
recent direct searches restrict severely the nature of the coupling of the dark matter to the $Z$ boson and set a limit $m_\chi \gtrsim 200$ GeV
(except in a very narrow region around the $Z-$pole region).
Using complementarity between spin dependent, spin independent and FERMI limits, we predict the nature of this coupling, more specifically
the axial/vectorial ratio that respects a thermal dark matter coupled through a $Z$-portal while not being excluded by the current observations. 
We also show that the next generation of experiments of the type LZ  or XENON1T will test
Z-portal scenario for dark matter mass up to 2 TeV. The condition of a thermal dark matter naturally predicts the spin--dependent scattering cross
section on the neutron to be $\sigma^{SD}_{\chi n} \simeq 10^{-40} \mrm{cm^2}$, which then becomes a clear prediction of the model and a signature testable 
in the near future experiments.
\end{abstract}

\maketitle


\maketitle


\setcounter{equation}{0}



\section{Introduction}
\label{Sec:introduction}

\noindent
The nature of the dark matter is one of the greatest puzzles in current science, once the dark matter constitutes
approximately 23\% of the Universe budget. There are many dark matter candidates in the literature, but the most seemingly
promising ones are the so--called WIMPs (Weakly Interacting Massive Particles) for having a thermal cross section at the electroweak
scale, naturally addressing the structure formation process, and being predicted in several extensions of the Standard Model (SM).
However, the key point of any extension Beyond the Standard Model (BSM) is to understand the mechanism lying behind the processes, 
namely annihilations, scatterings and possibly decays, which mantain the dark matter into thermal and kinetic equilibrium in 
the early stages of the history of the Universe, prior to its decoupling from the primordial thermal bath.
These same processes may be also active nowadays and can be probed by direct and indirect detection experiments. A microscopic approach, contrary to the effective one \cite{EFT}, requires the presence of a mediator between the dark and visible sectors : any (non)observation in DM experiments can then be used to restrict the mass and
coupling of the dark matter to its mediator given the condition it is thermal. Several particles were used as a mediator in the literature,
from the Higgs boson \cite{higgsportal,higgsportal2} to the supersymmetric pseudo scalar $A$ or scalar $H$ \cite{AHportal} passing through
dilaton \cite{dilatonportal} or $Z'$ in gauge extensions of the SM \cite{Zprime,Zprimecollider,Zprimeother}. Recently have been as well considered, alternatively, simplified setups without explicitly identifiying the mediator with a specific particle, see e.g. \cite{Buchmueller:2014yoa,simplified} .
\noindent
The $Z-$portal can be considered, like the Higgs portal, as a minimal 
extension of the Standard Model. Indeed the $Z$ is with the Higgs boson the only particles naturally present in the Standard Model 
which are able to play the role of mediator between the visible and the dark sector. However, the last LUX 
results\cite{Akerib:2013tjd}, combined with the invisible width of the Higgs excluded the Higgs-portal scenario for dark matter mass below 200 GeV \cite{higgsportal}.

 \noindent
 Even if the $Z-$portal scenario is $naturally$ present in a large number of extensions of the SM (sneutrino or 
 higgsino dark matter are $Z-$portal types as well as heavy neutrinos or models involving  kinetic mixing or dark photons). In general they are excluded
 in their minimal version because of their strong vectorial coupling necessary to respect relic abundance bounds. Such large vectorial couplings
  are indeed prohibited by direct detection limits \cite{Falk:1994es}. 
  In all these extensions, the axial coupling $A_\chi$ (see eq.(\ref{eq:Zlagrangian})) of the $Z$ boson to the dark matter is 
 naturally of the  order of magnitude of its vectorial coupling $V_\chi$. The deep reason is that in a framework of $SU(2)_L \times U(1)$ breaking
  the original $SU(2)_L$ condition ($V_\chi=A_\chi$) is only mildly modified by the dynamic of the breaking. 
The main idea of our work, 
is to generalize the $Z-$portal scenario without imposing any relation between the vectorial and axial part 
of the coupling of the $Z-$boson to the dark matter particle. We then study the 
nature of the coupling (vectorial versus axial) still allowed by the combined analysis of nowadays experiments.     
It has been shown recently in \cite{Lebedev:2014bba} and \cite{deSimone:2014pda} that the presence of $pure$ axial coupling 
 can reopen a large region of parameter space excluded by vectorial interactions, and we show that this is also the case in the $Z-$portal
 scenario. A Majorana dark matter is a typical example of a particle coupling purely axially to the $Z$ and to which our study directly applies.
  
 \noindent
 The paper is organized as follows. After a brief description of the model and our notation, we compute the annihilation cross section in a generic
 $Z-$portal dark matter scenario. We then apply the LEP constraints on the $Z-$width, spin-dependent/independent direct detection cross section, obtained by 
the LUX collaboration, and
 indirect detection from the FERMI telescope. We then explore the possibility that the $Z-$portal scenario can explain the excess of gamma-ray
 observed from the Galactic Center before concluding.  

\noindent
We give all the necessary analytical formulae in the appendix, whereas approximated ones 
 are used throughout the text to understand the physical phenomena at play. Our numerical analysis is obviously made by using the exact equations.

\section{The model}
\label{sec:model}

\subsection{The Lagrangian}

\noindent
The most generic way to describe the effective interaction of a dark matter particle with the Standard Model sector is
to write the couplings in term of its vectorial and axial part. Indeed the nature of the 
interaction determines completely the phenomenology of the dark sector. 
We then consider the following Lagrangian:

\begin{align}
\label{eq:Zlagrangian}
& \mathcal{L}=\frac{g}{4 \cos\theta_W}\left( \overline{\chi}\gamma^\mu \left(V_\chi-A_\chi \gamma^5\right) \chi Z_\mu+\right.\nonumber\\
& \left. \overline{f}\gamma^\mu \left(V_f-A_f \gamma^5\right) f Z_\mu\right)
\end{align}

\noindent
with $g$ the electroweak coupling\footnote{Notice that we have extracted the gauge coupling from the definition of $V_i$ and $A_i$. Indeed,
we can suppose that in an ultraviolet completion of the model (GUT--like framework), the gauge coupling should be naturally of the order of the 
electroweak coupling. In the literature, some authors define $g_A^i = g \times A_i$.
} ($g \simeq 0.65$), $V_{f,\chi}$ and $A_{f,\chi}$  the vectorial and axial charges respectively. $f$ represents the
Standard Model fermions with:

\be
V_f = 2\left(-2 q_f \sin^2 \theta_W +  T^3_f\right)   ~;~~~A_f= 2 T^3_f,
\ee

\noindent
$\theta_W$ being the Weinberg angle and $T^3_f$ the isospin number of the fermion $f$ with electric charge $q_f$.

\noindent
Notice that this lagrangian is not manifestly $SU(2) \times U(1)$ invariant and should be then complemented at high energy. A very simple option consists into higher dimensional operators coupling 
the DM with the Higgs boson and its covariant derivatives \cite{Cotta:2012nj,deSimone:2014pda} which give eq.(\ref{eq:Zlagrangian}) when the Higgs boson aquires {\it vev}. Alternatively one can consider a kinetic mixing scenario with the $Z^{'}$ lying at a much higher scale than the $Z$-boson. We will discuss this point in more details later on.

\subsection{Dark matter annihilation and PLANCK constraints}

\noindent
The dark matter phenomenology depends strongly on the different final states kinematically open.
For instance, helicity suppressions present in the case of two fermionic final states is absent for $ZZ$ or
$Zh$ final states. We have then distinguished three scenarios: $m_\chi < m_W$, $m_W < m_\chi < (m_Z + m_h)/2$
and $(m_Z + m_h)/2 < m_\chi$.

\subsubsection{\underline{$m_\chi < m_W$}}

\noindent
In this case the dominant channel is the dark matter annihilation into fermion pairs mediated by the $Z$ boson (see fig.(\ref{Fig:feynman})).
The reader can find in the appendix the general formulae for the annihilation cross section (eq.(\ref{Eq:sigvff1}) and (\ref{Eq:sigvff2})).
For a Dirac dark matter one obtains\footnote{We obviously run the numerical analysis with the exact formulae 
for the annihilation cross sections, the simplified equations are given to understand the dominant mechanisms dominating the process. Our results have been 
as well validated through the package MICROMEGAS \cite{Belanger:2013oya} (the authors want to thank particularly A. Pukhov for the support provided).}:

\bea
&&
\langle \sigma v \rangle_{f \bar f} \simeq \frac{g^4 m_\chi^2}{32 \pi \cos^2\theta_W m_Z^4}\nonumber\\
&&
 \sum_f n_c^f \left( |V_f|^2 + |A_f|^2  \right) {\left(1-\frac{4 m_\chi^2}{m_Z^2}\right)}^{-2}
\biggl[ 2 |V_\chi|^2 + \nonumber\\
&& |A_\chi|^2 \left( \frac{m_b^2}{m_\chi^2}  \frac{|A_b|^2}{\sum_f ( |V_f|^2 + |A_f|^2 ) } + \frac{v^2}{6} \right) 
\biggr]
\label{Eq:sigvff}
\eea

\noindent
$v$ being the relative (M\"{o}ller) velocity between annihilating dark matter particles ($ v \simeq 0.24$ at the decoupling time)
and $n_c^f$ the color number of the fermion $f$.
We should made several useful comments on this annihilation expression. First of all we remark that the s-wave contribution associated to the 
axial coupling $A_\chi$ is helicity suppressed and then it is not totally negligible only for the $b \bar b$ final state, although 
suppressed as $m_b^2/m_\chi^2$; the contribution to the annihilation cross-section from the axial coupling is then substantially dominated by 
the velocity dependent term at the decoupling time.

\begin{figure}
    \begin{center}
    \includegraphics[width=1.3in]{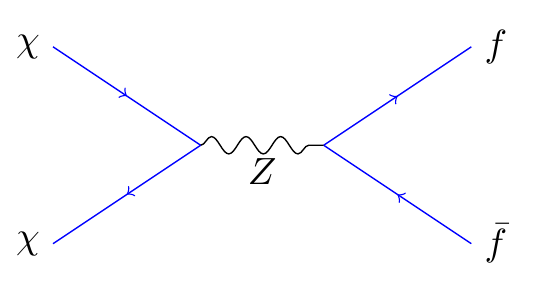}
    \includegraphics[width=1.3in]{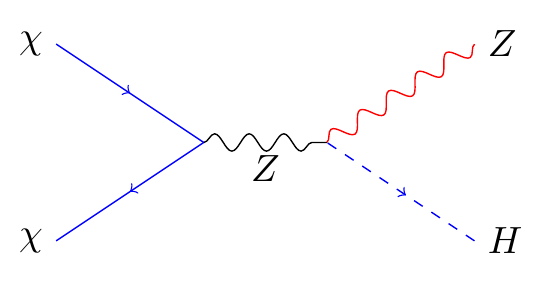}
    \includegraphics[width=1.3in]{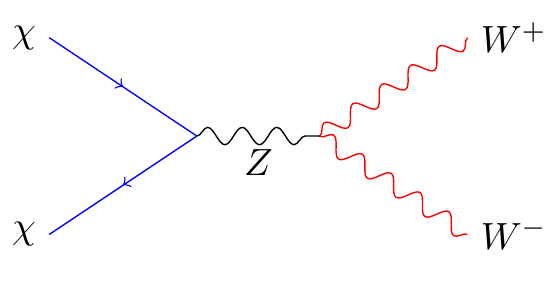}
    \includegraphics[width=1.3in]{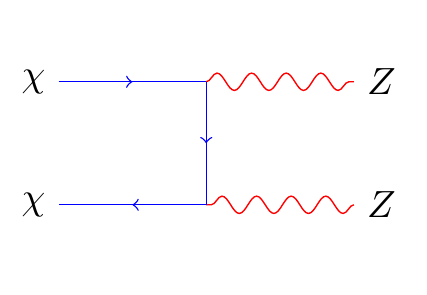}
              \caption{{\footnotesize
        Main Feynman diagrams contributing to the annihilation of dark matter in the $Z-$portal model. The dominant process are the 
        $b \bar b$ final state for $m_\chi \lesssim 100$ GeV and $Zh$ process for $m_\chi \gtrsim 100$ GeV}}
\label{Fig:feynman}
\end{center}
\end{figure}

\noindent
On the other hand, in presence of a sizable vector coupling $V_\chi$, the DM annihilation into fermions is rather effective and could be potentially the source of an indirect detection signal at present time. However, as discussed below, this possibility is already excluded by the very severe limits, from 
direct detection experiments, on the dark matter spin independent cross-section, obtained by LUX, which is very sensitive to $V_\chi$. 

\noindent
In the case of a Majorana dark matter the vectorial coupling cancels\footnote{in the case of a Majorana
fermion $\chi^M$, $\overline{\chi^M} \gamma^\mu \chi^M + h.c. = 0 $}, strongly relaxing the limits from direct detection, 
but implying, at the same time, poor detection prospects for indirect detection, 
because of the velocity and mass suppression of the DM annihilation cross-section. 
A Majorana dark matter in the $Z-$portal framework is then very difficult to detect, except in accelerators searches, where the nature
of the coupling does not play any role. On the other hand, to respect relic abundance, the model needs a relatively large
axial coupling $A_\chi$ to compensate the helicity suppression and avoid overabundance of dark matter. This situation is similar to a fermonic dark matter in the classical
Higgs portal where velocity suppression in $\langle \sigma v \rangle$ implies a large coupling to the Higgs, that is already excluded by spin--independent 
direct detection constraints as it has been shown in \cite{higgsportal}. Majorana dark matter however is not yet excluded 
by present direct detection constraints, as the spin--dependent bounds (depending only on the axial coupling) 
are much weaker than the spin--independent ones (which are
proportional to the atomic weight square $A^2$ of the nucleus target). 

\noindent
However, nothing forbids a Dirac dark matter to develop a hierarchy between its vectorial $V_\chi$ and axial $A_\chi$ coupling, opening
a parameter space where the direct, indirect accelerators and cosmological constraints are still satisfied. Indeed, the pure vectorial, or
well tempered mixed coupling to the $Z$ is excluded since long time ago (see \cite{Falk:1994es} for instance) since
the spin-independent direct detection cross section for a $Z-$portal with $A_\chi \simeq V_\chi$ respecting WMAP/PLANCK constraints \cite{Hinshaw:2012aka,Ade:2013zuv} 
exceeds the present limit obtained by LUX \cite{Akerib:2013tjd} or XENON100 \cite{Aprile:2012nq} collaborations.

\noindent
As an illustration, we show in fig.(\ref{Fig:alpha1}) in full line the constraint on $A_\chi$ for $A_\chi = V_\chi$, obtained by LUX, as function of the 
mass of the dark matter. From eq.(\ref{Eq:sigvff}) we understand that when
$m_\chi \lesssim m_Z$ the relic abundance constraint $\langle \sigma v \rangle = \langle \sigma v \rangle_{\mrm{thermal}} \simeq 2.6 \times 10^{-9} ~ \mrm{GeV^{-2}}$ implies that
$m_\chi \times A_\chi \simeq$ const, except the $Z$-pole region $m_\chi \simeq m_Z/2$ when the value of $A_\chi$ should be particularly small since the 
cross-section encounters a s-channel resonance\footnote{In the resonance region the velocity expansion, used for our analytical 
estimates, is not valid. As mentioned our result are based on the numerical evaluation of the full expression of the annihilation cross-section, 
including the $Z$-width in the propagator. For an analytical description of the resonance region one could refer e.g. to \cite{Griest:1990kh,Gondolo:1990dk}.}.

\noindent
For completeness, we also plotted in fig.(\ref{Fig:all}) the PLANCK constraints for different values of the ratio $\alpha = |A_\chi| / |V_\chi|$. 
An interesting example is the pure axial case, because less constrained by direct detection as we will see in the next section. In
this case, the dominant term in the cross section is\footnote{One can notice that a similar pure axial analysis in the case of an extra $Z'$ was
published in \cite{Lebedev:2014bba}.}

\bea
&&
\langle \sigma v \rangle_{\mrm{axial}} \simeq \frac{g^4 m_\chi^2 v^2 |A_\chi|^2 \sum_f n_c^f \left( |V_f|^2 + |A_f|^2  \right)
}{96 \pi \cos^2 \theta_W m_Z^4} 
\nonumber
\\
&&
\times {\left(1-\frac{4 m_\chi^2}{m_Z^2}\right)}^{-2} \nonumber\\
&&
= 3.2 \times 10^{-8}~\mrm{GeV^{-2}} |A_\chi|^2 {\left(\frac{m_\chi}{m_Z}\right)}^2 {\left(1-\frac{4 m_\chi^2}{m_Z^2}\right)}^{-2}
\nonumber
\label{eq:sigmaff_an}
\eea

\noindent
and gives the cosmologically favored value $\langle \sigma v \rangle_{\rm PLANCK}\simeq 2.6 \times 10^{-9}\,{\mbox{GeV}}^{-2}$, for e.g. $m_\chi =9\,\mbox{GeV}$ and 
$|A_\chi| \sim 3$. As can be seen, eq.(\ref{eq:sigmaff_an}) is in agreement with 
the numerical result we obtained in fig.(\ref{Fig:all}) for the axial case $V_\chi=0.01 A_\chi$.
In order to respect the relic abundance for a given mass, the value of the axial coupling should be much larger, order of the velocity suppression $\sqrt{v^2/6} \simeq 0.1$, than the one required for the vectorial coupling $V_\chi$ to satisfy the same constraint. This can also be observed on fig.(\ref{Fig:all}) by comparing the pure
axial ($|A_\chi | \gg |V_\chi|$) and the mixed case ($|A_\chi| = |V_\chi|$).

\subsubsection{\underline{$m_W < m_\chi < (m_Z + m_h)/2$}}

\noindent
In this small (20 GeV) window, the $b \bar b$ s-channel final state competes with the $ZZ$ t-channel and
the $WW$ s-channel final states (see fig.(\ref{Fig:feynman})). For $m_\chi > m_Z$ one obtains\footnote{The units will be $\mrm{[GeV^{-2}]}$ for all the observable
thorough the paper, except if specified.}
(we let the reader to have a look at the appendix for the exact analytical 
formulae): 

\bea
&&
\langle \sigma v \rangle_{ZZ} \simeq \frac{g^4}{16 \pi \cos^2 \theta_W M_Z^2}\nonumber\\
&&
\left(|A_\chi|^2 |V_\chi|^2+\frac{v^2}{3} |A_\chi|^4 {\left(\frac{m_\chi^2}{m_Z^2}\right)}\right)
\nonumber
\\
&&
\simeq 5.4 \times 10^{-7} |A_\chi|^2 \left(|V_\chi|^2+0.05 |A_\chi|^2 {\left(\frac{m_\chi^2}{m_Z^2}\right)}\right)
\nonumber
\\
&&
\langle \sigma v\rangle_{WW} \simeq \frac{g^4 \tan \theta_W }{16 \pi M_W^2}\left(|V_\chi|^2 \left(1-\frac{v^2}{6}\right)\right.
\nonumber\\
\\
&&
\left. +|A_\chi|^2 \frac{v^2}{3}\right) {\left(\frac{m_\chi^2}{m_W^2}\right)}
\nonumber
\\
&&
\simeq 1.8 \times 10^{-7} |V_\chi|^2 {\left(\frac{m_\chi^2}{m_W^2}\right)}\,[{\mbox{GeV}}^{-2}]
\eea

\noindent
for $v \sim 0.24$, typical value of the M\"{o}ller velocity at decoupling.
However, the $ZZ$ channel is never the dominant one, and the $WW$ final states dominates largely this window region. Notice the absence of the axial coupling 
$A_\chi$ in the leading term of the cross-section of the $W^+ W^-$ final state. Indeed, by angular momentum and CP conservation arguments, only vectorial operators can contribute to the s-wave term. 
One can also remark that both cross-sections feature (although only in the p-wave term for the $ZZ$ channel) enhancement factors $\frac{m_\chi^2}{m_V^2},\,V=W,Z$. These originate from the longitudinal components of final state gauge bosons. We will further develop this point in the next subsection.

\subsubsection{\underline{$m_\chi > (m_Z + m_h)/2$}}

\begin{figure}
    \begin{center}
    \includegraphics[width=3.in]{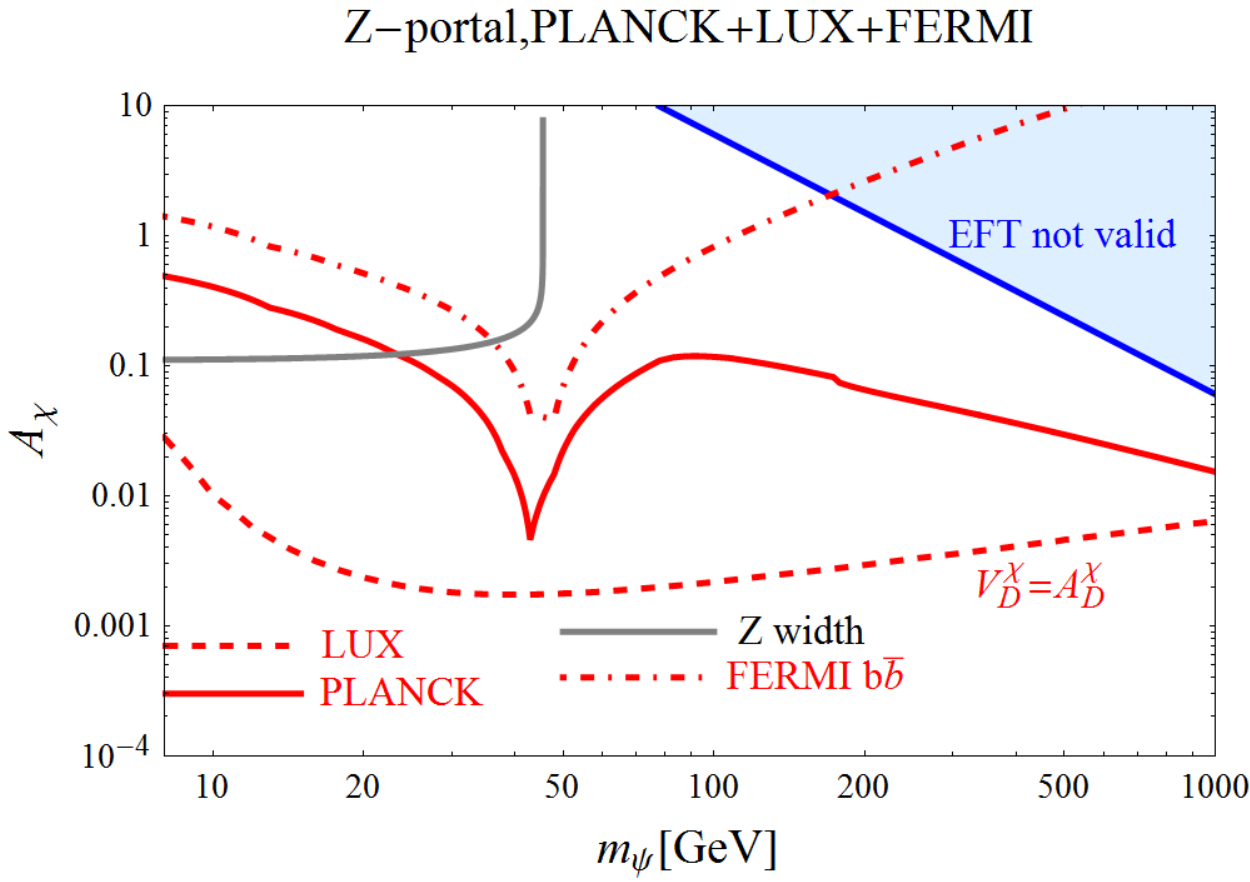}
              \caption{{\footnotesize
        PLANCK, $Z-$width, LUX and FERMI constraint i  the plane ($m_\chi$, $A_\chi$) in the case $\alpha = A_\chi / V_\chi=1$. 
        The lines corresponds  to the PLANCK constraint (full red) ) \cite{Ade:2013zuv}, $Z-$width (full grey) \cite{Agashe:2014kda}, FERMI dwarf galaxies~(dash-dotted red) \cite{Ackermann:2013yva} and LUX (dashed red) \cite{Akerib:2013tjd}. In the blue 
				region the lagrangian~(\ref{eq:Zlagrangian}), assuming that it arises from the operator introduced in \cite{deSimone:2014pda}, 
				does not provide a consistent 
				description and should be replaced by the UV complete model.}}
\label{Fig:alpha1}
\end{center}
\end{figure}

\noindent
As soon as the $Z h$ final state is open, it largely dominates the dark matter annihilation process, in competition with the $W^+ W^-$ channel.
One can also notice, in figs.(\ref{Fig:alpha1}) and (\ref{Fig:all}), the little "bump" for $m_\chi=175$ GeV corresponding to the opening of the $t \overline t$ final state (contributing
to around 10\% to the total annihilation process): one needs a lower value of $|A_\chi|$ once the $t \overline t$ final state is open 
as this new contributions increase the dark matter depletion at the decoupling time.  The channels $Zh$ and $t \bar t$ are on 
the proportion 90\% / 10 \% whereas the $WW$ channel depends strongly on $|V_\chi |$ and thus on $\alpha$. We obtain in the limit
$m_\chi > m_Z$

\bea
&&
\langle \sigma v \rangle_{Zh} \simeq 2.4 \times 10^{-8} \left( \frac{m_\chi^2}{m_Z^2} \right)  |A_\chi|^2 ~~\mrm{[GeV^{-2}]}
\nonumber
\\
&&
\langle \sigma v \rangle_{t \bar t} \simeq 1.7  \times 10^{-7}  |A_\chi|^2~~\mrm{[GeV^{-2}]}
\label{Eq:sigv3}
\eea

\noindent
We remark that one should put particular care in interpreting these results in the high DM mass region. By increasing the DM mass, indeed, the 
DM annihilation cross-section starts to be sensitive to the UV completion leading to eq.(\ref{eq:Zlagrangian}) and its behavior might be different from 
the one reported. We remind in particular that the enhancement factor $m_\chi^2/m_V^2$ may led to violation of unitarity unless a UV complete model, to describe consistently the annihilation into gauge bosons at high DM masses, is adopted. For illustration we show in fig.(\ref{Fig:alpha1}), in the light blue region, the range of values of 
$A_\chi$ for which the lagrangian (\ref{eq:Zlagrangian}) does not provide a reliable description, in the case it originates from the effective operator reported in section III E with the scale of New Physics $\Lambda$ set to 
be of the order of the EW scale \cite{deSimone:2014pda}. For the considered range of DM masses, analyzed in this work, our results can be regaded as reliable.

\section{Analysis}

\subsection{$Z-$width constraint}
\label{Sec:zwidth}

\noindent
As the invisible Higgs constraint is the strongest one in the case of Higgs portal, one can notice that the $Z-$width constraint
is the strongest one in the $Z$-portal when $m_\chi < m_Z/2$. Indeed,the total width is given by 
$\Gamma_Z= \Gamma_Z^{\mrm{SM}} + \Gamma^{\mrm{\chi}}_Z$,
with

\bea
&&
\Gamma_Z^{\mrm{\chi}}= \frac{g^2}{192 \pi \cos^2 \theta_W} m_Z \sqrt{1 - \frac{4 m_\chi^2}{m_Z^2}} \times
\label{Eq:zwidth}
\\
&&
\left[|V_\chi|^2  \left( 1 + \frac{2 m_\chi^2}{m_Z^2} \right) + |A_\chi|^2 \left( 1 - \frac{4 m_\chi^2}{m_Z^2} \right) \right]
\nonumber
\eea

\begin{figure}
    \begin{center}
    \includegraphics[width=3.in]{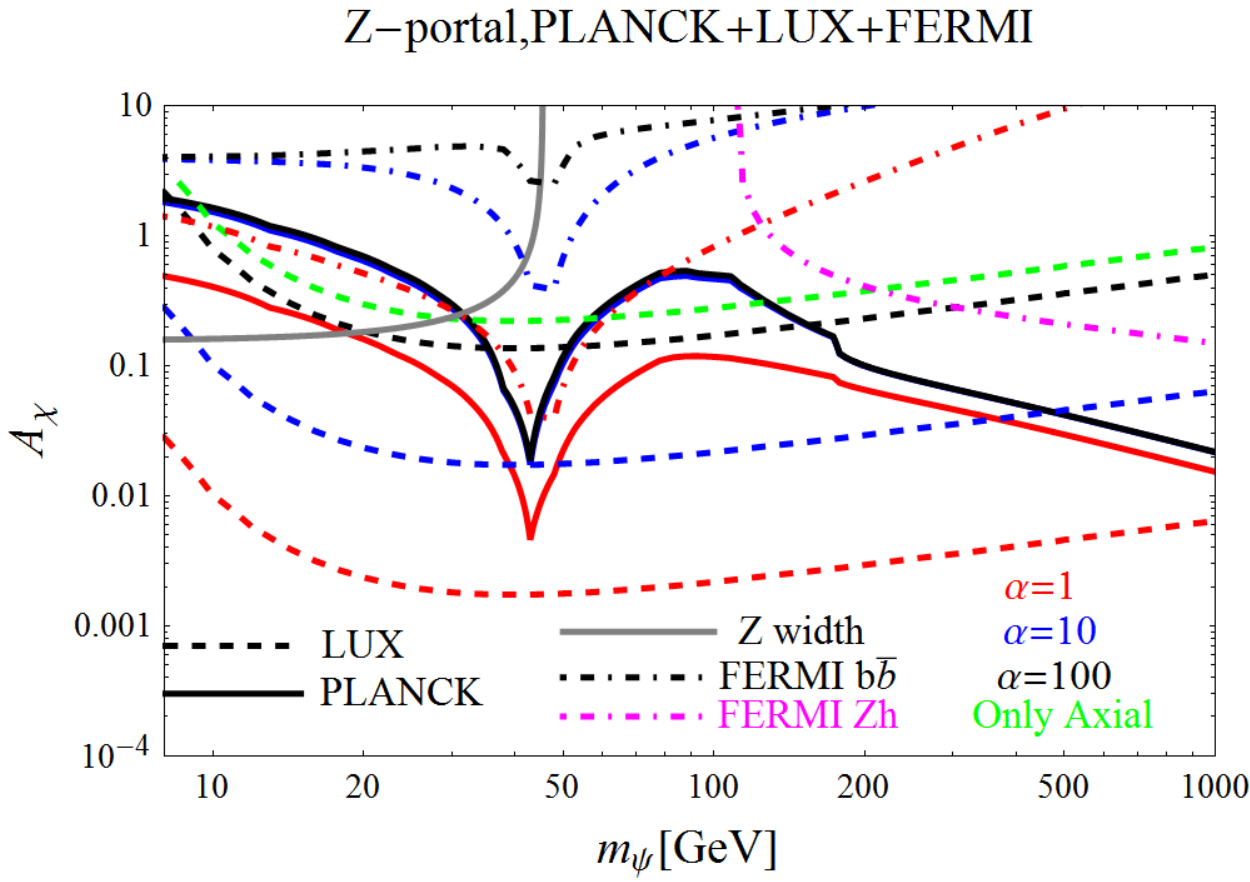}
              \caption{{\footnotesize
        Summary of all the constraints analyzed in this work, for different values of the ratio $\alpha = |A_\chi|/|V_\chi|$. The full lines
        correspond to the PLANCK constraint, the dashed line to LUX constraint (the green one for the pure axial case) whereas
        the dot-dashed ones are the FERMI limits from dwarf galaxies. See the text for details.}}
\label{Fig:all}
\end{center}
\end{figure}

\noindent
Imposing that the $Z-$width lies within the LEP experimental bounds \cite{Agashe:2014kda}   
$\Gamma_Z= 2.4952 \pm 0.0023$ GeV, gives a limit on the non--standard invisible width, 
$\Gamma_Z^{\mrm{\chi}} \lesssim 2.3$ MeV. We illustrate this constraint 
in the specific case  $|A_\chi| = |V_\chi|$ on fig.(\ref{Fig:alpha1}) (solid gray line).
It is easy to compute from eq.(\ref{Eq:zwidth}) the relation between $|A_\chi|$ and $|V_\chi|$ induced by the condition 
$\Gamma_Z^{\chi}  \lesssim 2.3$ MeV: $|V_\chi|^2 + |A_\chi|^2 = \frac{192 \pi \cos^2\theta_W}{g^2} \frac{\Gamma^{\chi}_Z}{m_Z}
\lesssim  0.03$. This limit is independent on the dark matter mass, even near the threshold.
For a more complete analysis, with different values of the ratio $\alpha = A_\chi / V_\chi$, one can have a look at fig.(\ref{Fig:all}): it is interesting to notice that, contrarily to direct and indirect detection cases, the $Z-$width is almost insensitive to the nature of the coupling
(see eq.(\ref{Eq:zwidth})): the axial and vectorial contributions are of the same order of magnitude.

\noindent
We can then directly link the relic abundance limit with the $Z$-width constraint. Combining eqs.(\ref{Eq:sigvff}) and (\ref{Eq:zwidth}) gives:

\bea
&&
\frac{\langle \sigma v \rangle}{\Gamma_Z^{\chi}}= \frac{12 g^2  m_\chi^2 \sum_f n_c^f \left( |V_f|^2 + |A_f|^2  \right)  }{m_Z^5 (1 + \alpha^2)} \times
\nonumber
\\
&&
\biggl[ 2 + \alpha^2 \left(\frac{b^2}{m_\chi^2} + \frac{v^2}{6}\right)   \biggr] {\left(1-\frac{4 m_{\chi}^2}{m_Z^2}\right)}^{-5/2}
\eea

\noindent
with $b^2= \frac{m_b^2 |A_b|^2}{\sum_f \left( |V_f|^2 + |A_f|^2  \right)}$ . We can then compute the minimum value of $m_\chi$
allowed by both the relic abundance constraint of not over close the Universe ($\langle \sigma v \rangle \geq 2.6 \times 10^{-9}~\mrm{GeV^{-2}}$)
and the $Z-$width to stay in the experimental limit set by LEP ($\Gamma_Z^{\chi} < 2.3$ MeV). In the pure vectorial case ($\alpha=0$), 
one obtains $m_\chi \gtrsim \frac{10^{-6} m_Z^5}{12 g^2 \sum_f \left( |V_f|^2 + |A_f|^2  \right)} \simeq 23$ GeV, whereas the pure axial case
($\alpha \gg 1$) gives $m_\chi \gtrsim 31.5$ GeV. We can recover these values with a more complete numerical analysis illustrated in figs.(\ref{Fig:alpha1})
and~(\ref{Fig:all}). In the case of Majorana DM we have $V_\chi=0$ and the decay width of the $Z$ into DM is reduced by a symmetry factor $1/2$. This leads to 
a slightly weaker limit of 28.5 GeV.

\subsection{Direct detection cross section}

\noindent
Recently, the LUX collaboration \cite{Akerib:2013tjd} set the strongest constraints on the spin independent (SI) as well as spin dependent (SD)
dark matter-nucleon cross section. Other experiments are still in the race like XENON100  \cite{Aprile:2012nq}
and PandaX \cite{Xiao:2014xyn} which will soon provide
their new limits. The interesting point in direct detection constraints is the complementarity between the SD and 
SI searches. Indeed, contrary to the conventional Higgs portal scenario, where the scalar interaction 
contributes only to the SI scattering cross-section, in the $Z$-portal framework one can use a combined analysis to restrict the ($V_\chi$,$A_\chi$) parameter space for each value of the 
dark matter mass: the SI elastic scattering cross section on a nucleon N, $\sigma_{\chi N}^{SI}$ being exclusively dependent\footnote{see 
the appendix for the complete set of analytical formulae} on $V_\chi$ whereas
the SD elastic scattering being exclusively dependent on $A_\chi$.
 It becomes then possible to express the thermal cross section $\langle \sigma v \rangle$
exclusively as function of the $physical$ $observables$ $\sigma^{SI}_{\chi N}$ and $\sigma^{SD}_{\chi N}$ and the mass of the dark matter
particle as was done in the case of a generic $Z'$ model in \cite{Arcadi:2013qia}. It is easy to understand that the expression is of the form
$\langle \sigma v \rangle \simeq c_1 \times \sigma^{SI}_{\chi N} + c_2 \times \sigma^{SD}_{\chi N}$, with $c_1$ and $c_2$ of the same order of magnitude.
The limit on the spin independent cross section settled by LUX experiment \cite{Akerib:2013tjd} being so stringent (almost 4 to 5 orders of magnitudes
compared to the spin dependent one, also set by LUX) that the 
conservative upper limit on $\langle \sigma v \rangle$ is uniquely determined by
the upper limit on
$\sigma^{SD}_{\chi N}$, and more precisely $\sigma^{SD}_{\chi n}$ given by LUX\footnote{LUX, like other xenon-type detectors, is much more sensitive to the scattering cross section on neutrons since the contribution from neutrons to the nuclear spin exceeds by an order of magnitude the one of protons.} which is the strongest spin dependent constraint\footnote{An analysis
on the spin dependent constraint from $\chi n$ interaction was already made in 2012 for the XENON experiment in \cite{Garny:2012it}.}
 at present \cite{Buchmueller:2014yoa,Malik:2014ggr}.

\noindent
We then obtain at the first order in velocity (see the appendix for the complete set of analytical formulae) in the case of $m_\chi \gtrsim 100$ GeV:

\bea
\langle \sigma v \rangle \simeq \langle \sigma v \rangle_{Zh} \simeq \frac{10^{-2}}{\alpha^{SD}_n} \left( \frac{m_\chi}{1~\mrm{GeV}} \right)^2 
\sigma^{SD}_{\chi n}
\label{Eq:sigmavsigmasd}
\eea

\noindent
where the coefficient $\alpha^{SD}_n$ is defined in the appendix and is given by:
\begin{align} 
& \alpha^{SD}_n =\sum_A \eta_A
 \biggl[ A_u(\Delta_u^p S_p^A 
+ \Delta_d^pS_n^A)\nonumber\\
& + A_d \left( (\Delta_d^p + \Delta_s^p) S_p^A+ (\Delta_u^p + \Delta_s^p)S_n^A  \right) \biggr]^2\nonumber\\ 
&\frac{1}{\sum_A \eta_A (S_p^A + S_n^A)^2}\frac{\left(A_u \Delta^p_u+A_d \left(\Delta^p_d+\Delta^p_s\right)\right)^2}{\left(A_u \Delta^n_u+A^d_D \left(\Delta^n_d+\Delta^n_s\right)\right)^2}
\end{align}

\noindent
The sum is over the xenon isotopes, with relative abundance $\eta_A$. In the case of xenon detector like LUX we have two contributions from ${\rm Xe}^{129}$ and ${\rm Xe}^{131}$ and $\alpha_n^{SD}\simeq 0.65$.  

\noindent
It is interesting to notice that, the lower bound on $\langle \sigma v \rangle$, from the requirement of non-overclosure of the universe by a thermal relic, 
turns into a lower bound on $\sigma^{SD}_{\chi n}$ which can be tested by future experiments.  
 Indeed, under the conservative hypothesis $\langle \sigma v \rangle \gtrsim 2.6 \times 10^{-9}~\mrm{GeV^{-2}}$ one obtains from
eq.(\ref{Eq:sigmavsigmasd})

\be
\sigma^{SD}_{\chi n} \gtrsim 1.2 \times 10^{-38} ~\alpha^{SD}_n \left( \frac{100~\mrm{GeV}}{m_\chi} \right)^2 ~\mrm{cm^2}
\ee

\begin{figure}
    \begin{center}
    \includegraphics[width=3.in]{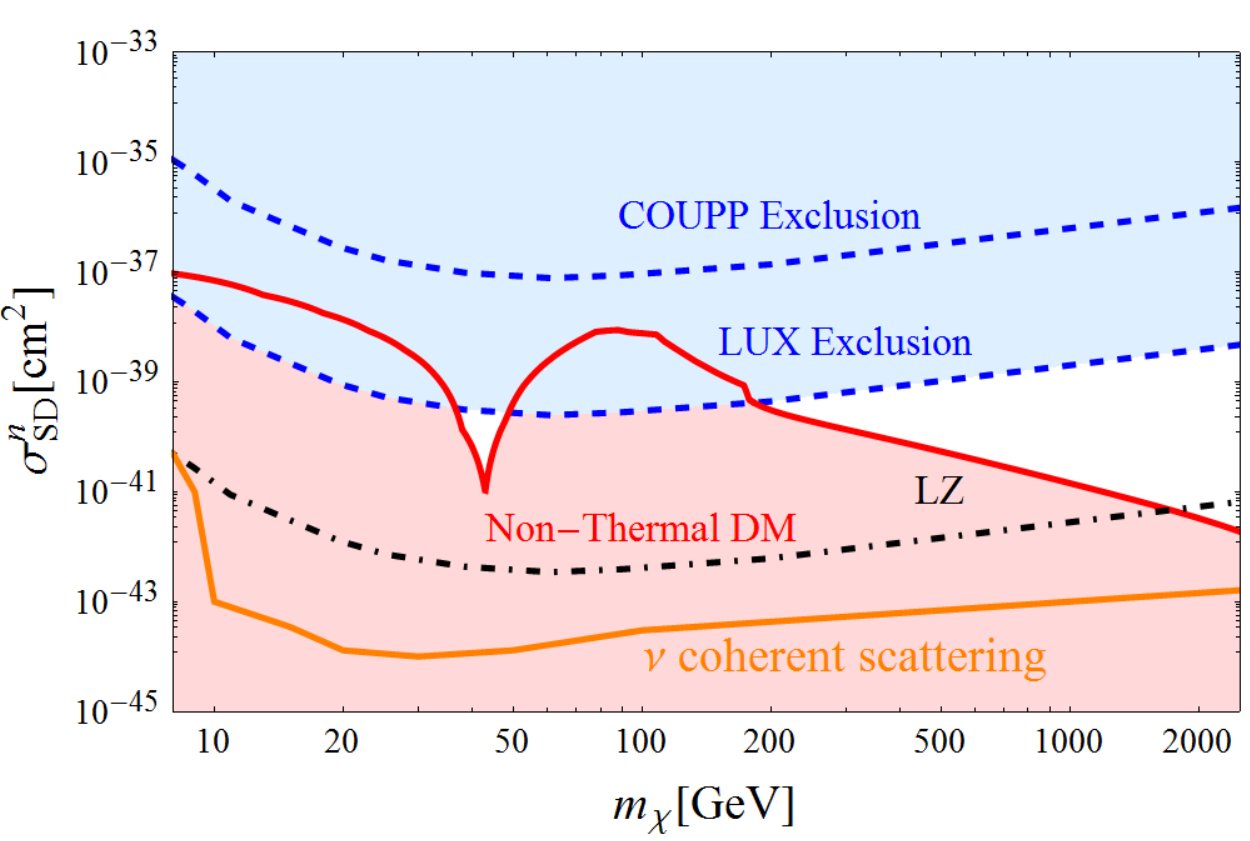}
              \caption{{\footnotesize
        Limit on the neutron-$\chi$ spin dependent cross section s function of $m_\chi$ and prospect for the future LZ project. We also present
        the neutrino scattering limit \cite{Billard:2013qya} which is lying inside the region where dark matter should have a non-thermal history to avoid
        the overclosure of the Universe.}}
\label{Fig:sigmasd}
\end{center}
\end{figure}

\noindent
Our purpose is clearly illustrated in fig.(\ref{Fig:sigmasd}) where we have plotted the spin dependent scattering cross section of the dark matter
on the neutron as function of the dark matter mass in comparison with limits of COUPP \cite{Behnke:2012ys,DelNobile:2013sia} and 
LUX \cite{Buchmueller:2014yoa} as well the expected sensitivity for the future LZ detector \cite{Buchmueller:2014yoa,Malling:2011va} and the determination,
provided in \cite{Buchmueller:2014yoa}, of the neutrino background \cite{Billard:2013qya}, which sets the maximal sensitivity achievable for this kind of 
direct dark matter searches. We notice that, except for a little region
around the $Z-$pole mass, the $Z-$portal model is excluded for dark matter mass below $\simeq$ 200 GeV, a situation comparable 
with the Higgs-portal model \cite{higgsportal}.

\begin{figure}[!]
    \begin{center}
    \includegraphics[width=3.in]{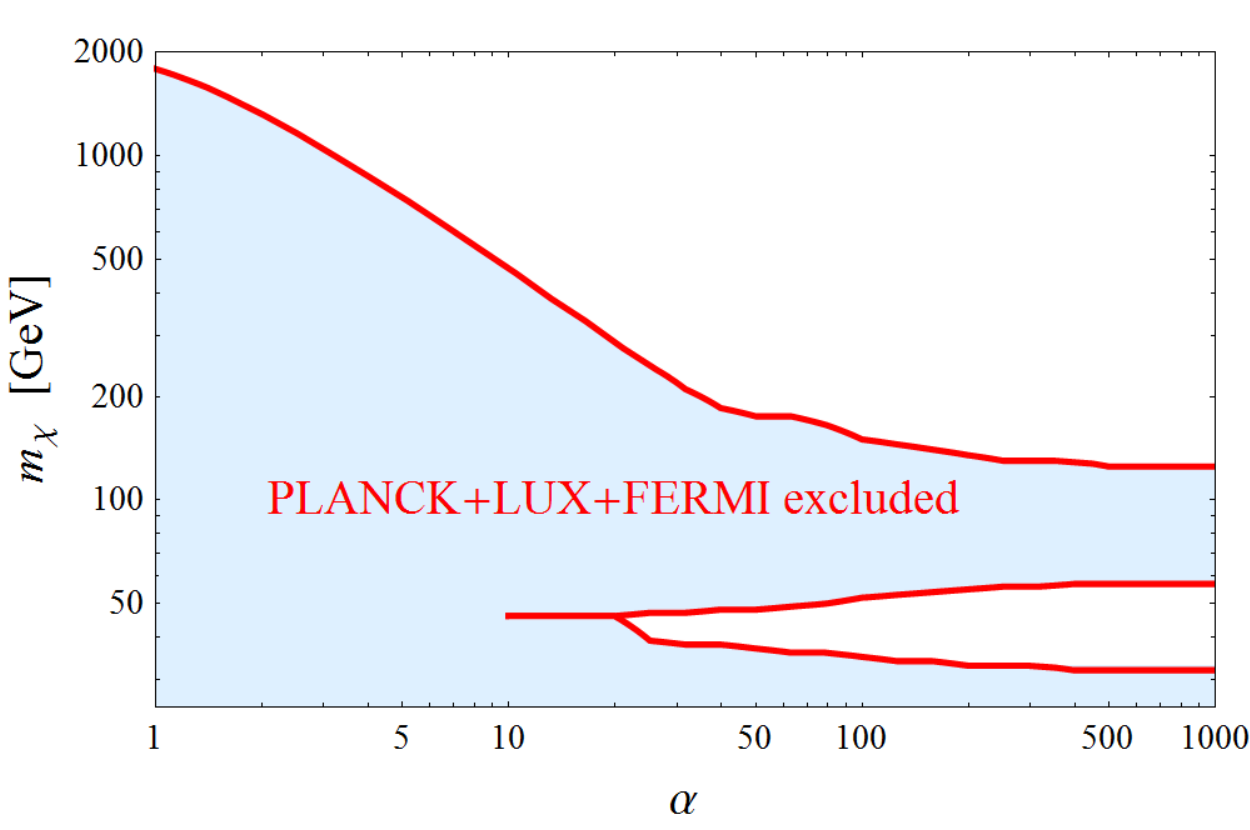}
              \caption{{\footnotesize
        Region allowed by the combined LUX/PLANCK/LEP/FERMI constraint in the plane ($\alpha = A_\chi/V_\chi, m_\chi$). We notice that, except
        in the $Z-$pole region, $m_\chi \gtrsim 130$ GeV for any value of $\alpha$, and even $m_\chi \gtrsim 2$ TeV for $\alpha=1$.}}
\label{Fig:alphamchi}
\end{center}
\end{figure}

\noindent
One can also better understand the situation by computing the ration $\alpha = A_\chi / V_\chi$ necessary to respect in the meantime the
LUX and PLANCK constraint, which is illustrated in fig.(\ref{Fig:alphamchi}). We clearly see that, for dark matter mass below 1 TeV,
$\alpha \gg 1$ which means that the coupling of the thermal dark matter to the $Z$ boson should be almost purely axial to respect both constraints. It is only for $m_\chi \gtrsim 2\,\mbox{TeV}$ that 
the vectorial nature of the dark matter begins to be allowed due to the weakness of spin-independent limit set by LUX for such heavy masses.

\noindent
Our determination of the limits from DM Direct Detection have been validated by complementing our analytical treatment with the numerical package 
described in \cite{DelNobile:2013sia}.

\subsection{FERMI constraint}

\noindent
Indirect detection of dark matter is also an efficient field to constraint extensions of the Standard Model. The most effective limits are at the moment given by
$\gamma$-ray emission in dwarf galaxies, which can provide very strong limits on the annihilation cross-section in view of the large dark matter / visible matter ratio in these objects. Limits on the DM annihilation cross-section into fermion and $W$-boson pairs are provided by the FERMI collaboration in \cite{Ackermann:2011wa,Ackermann:2013yva} while 
constraints in the other final states including gauge/higgs bosons have been determined e.g. \cite{Fedderke:2013pbc,Geringer-Sameth:2014qqa}. 

\noindent
We show in fig.(\ref{Fig:all}) the constraints on the axial coupling $A_\chi$ exctracted from the FERMI analysis, on the two channels dominating the DM pair annihilation cross-section at present times, i.e. the $b \bar b$, for $m_\chi \lesssim 100\,\mbox{GeV}$, and $Zh$, in the high mass regions,  for $|V_\chi|=|A_\chi|,0.1 |A_\chi|, 0.01 |A_\chi|$. We show as well the bounds from the correct relic density and from direct detection obtained by LUX. The bounds from FERMI are much weaker especially with respect to the ones extracted from LUX.

\begin{figure}
    \begin{center}
    \includegraphics[width=3.in]{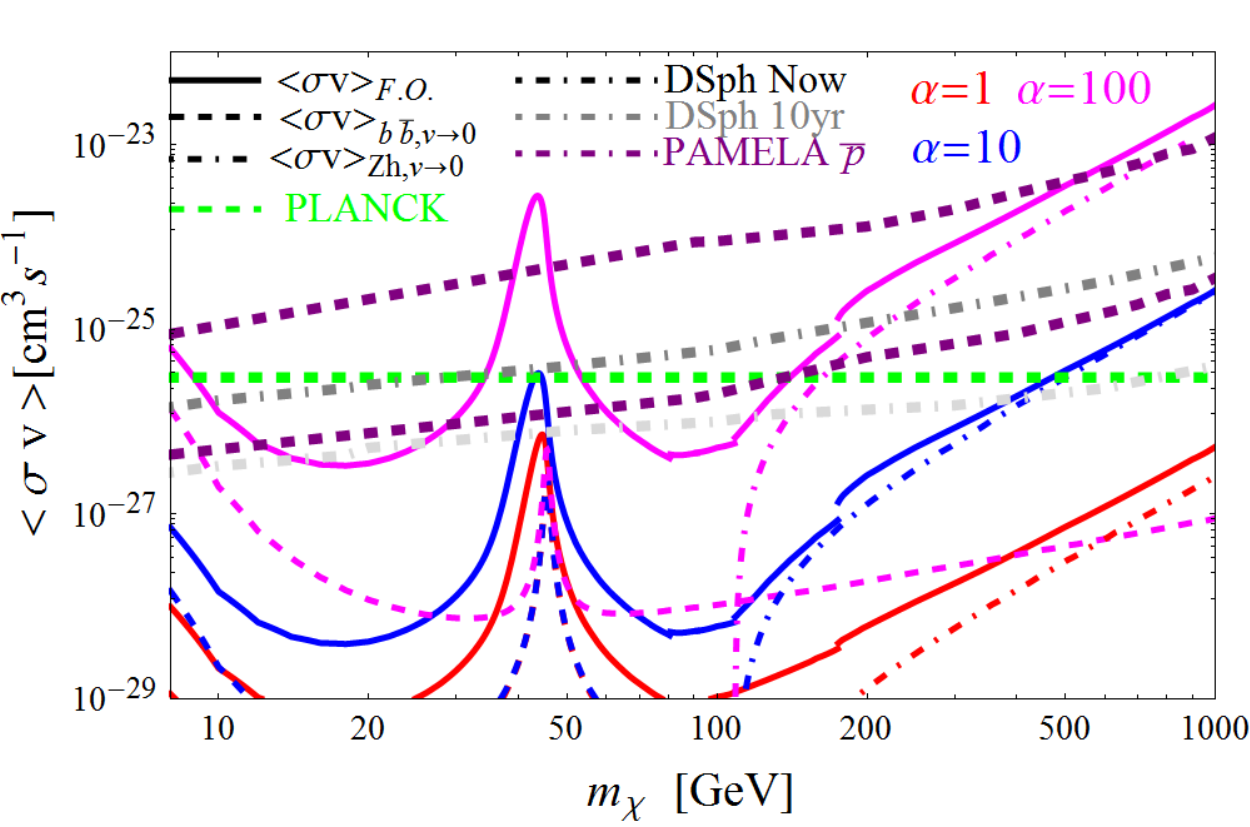}
              \caption{{\footnotesize
      Maximal value, as function of the Dark Matter mass, allowed by LUX of the total annihilation cross-section at the time of freeze-out (solid lines) and of the annihilation cross-section at present times in the $b\bar b$ (dashed lines) and $Zh$ (dot-dashed lines) final states, for $\alpha=\frac{|A_\chi|}{|V_\chi|}=1\,(\mbox{Red}), 10\,(\mbox{Blue}), 100\,(\mbox{Magenta})$. The gray dashed line represents the current FERMI constraint from observation of DSph, while the light-gray one is the projected sensitivity in 10 years. The two violet dashed lines are the limits from antiprotons extracted from PAMELA in \cite{Fornengo:2013xda}  for the ``MAX'' and ``MIN'' choices of the propagation parameters.}}
\label{Fig:fermi}
\end{center}
\end{figure}

\noindent
The impact of dark matter indirect detection, with respect to the other observables, is evidenced in fig.(\ref{Fig:fermi}). In this figure we plotted the dark matter annihilation cross-section at freeze-out, for three values of $\alpha=|A_\chi|/|V_\chi|=1,10,100$ while setting $V_\chi$ to the limit imposed by LUX, together with the present time annihilation cross-sections for the most relevant channels ($b \bar b$ and $Zh$). These values are compared with the present limits \cite{Ackermann:2013yva} form dwarf spherodial galaxies\footnote{These are actually conservative limits. New preliminary results from the FERMI collaboration \cite{FermiSymposium}, have set stronger constraints excluding dark matter candidates, with present time $b \bar b$ annihilation cross-section of the order of the thermal value, with mass below 100 GeV.} as well as a future projected sensitivity \cite{DphS10years}. We can see that the constraints from dark matter indirect detection are irrelevant, for dark matter masses below the $Zh$ thresold. Indeed the thermally averaged cross-section of the dark matter at freeze-out can achieve the cosmologically favored value only for $\alpha$ between, approximately, 10 and 100 (in the case $\alpha=1$ the $b \bar b$ cross-section at present time is actually of the same order of the one at freeze-out. However the annihilation cross-section for $\alpha=1$ must be much lower than the PLANCK favored value in order to satisfy the constraints on $V_\chi$ from LUX). For such values of $\alpha$, the $W^{+}W^{-}$, $ZZ$ and the $f \bar f$ final are strongly velocity dependent such that their present time annihilation cross-sections are suppressed with respect to the cosmological ones. The only sizable contribution is given by the $b \bar b$ channel which, in the $v \rightarrow 0$ limit, is determined by its helicity suppressed s-wave term. 

\noindent
The situation is instead different if the $Zh$ final state is kinematically accessible. In such a case, the annihilation cross-section is always dominated by an unsuppressed s-wave contribution and its present time value is not very different from the one relevant for the DM relic density. Current indirect detection constraints in such a case limit the rise of the cross-section with the DM mass, although the thermal value is still viable for $\alpha$ up to 100 and a future increase of the sensitivity can probe thermal DM candidates for $\alpha$ even lower than 10. For comparison we have also reported in fig.(\ref{Fig:fermi}) limits from antiprotons derived in \cite{Fornengo:2013xda} for the two extremal choices, dubbed ``MIN'' and ``MAX'', of the propagation parameters. The different choices of these parameters induce extremely strong uncertainties in the bounds. We nonetheless remark that for the ``MAX'' choice antiprotons bounds already probe the thermal DM region for $\alpha \sim 100$.    

\subsubsection*{Galactic Center signal?}

\noindent
Recently, the authors of \cite{Daylan:2014rsa,Hooper:2010mq}, reanalyzing the FERMI data from the Galactic Center, made the claim (also confirmed more recently by \cite{Calore:2014xka}) of an excess of gamma rays around the galactic center compatible a dark matter mass 
of $\simeq 30-50\,\mbox{GeV}$, mostly annihilating into $b \bar b$ with an annihilation cross section of $\langle \sigma v \rangle \simeq 2 \times 10^{-26} ~\mrm{cm^3 s^{-1}}$. 

\begin{figure}
    \begin{center}
    \includegraphics[width=3.in]{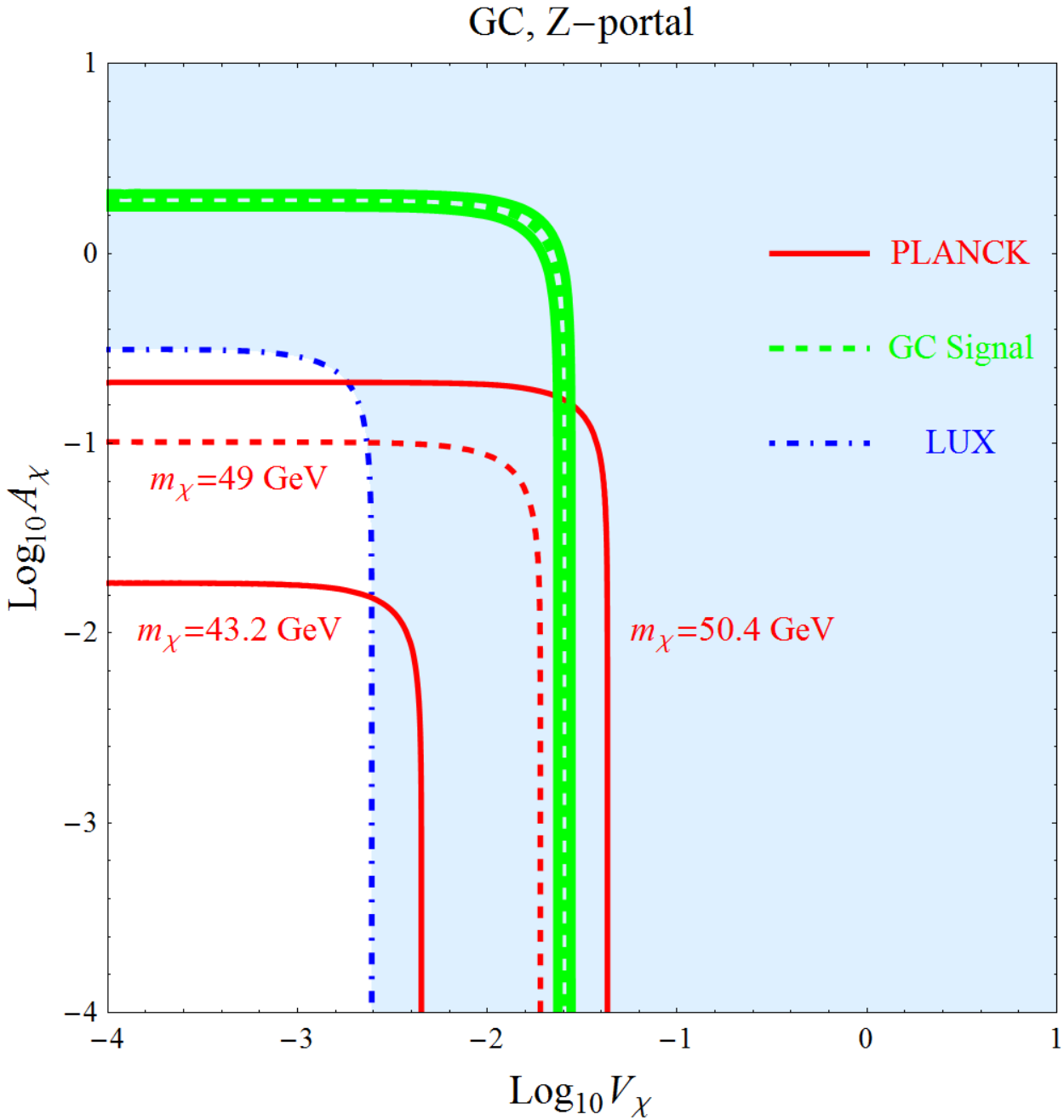}
              \caption{{\footnotesize
        Combined constraints in the plane ($V_\chi$, $A_\chi$) for values of $m_\chi$ and $\langle \sigma v \rangle$ as determined in \cite{Calore:2014xka}: $m_\chi \in [43.2,50.4]\,\mbox{GeV}$ and $\langle \sigma v \rangle_{v \rightarrow 0}(\bar \chi \chi \rightarrow b \bar b)=1.76 \pm 0.28 \times 10^{-26}\,{\mbox{cm}}^3{\mbox{s}}^{-1}$. The light-blue region is excluded by LUX.}}
\label{Fig:hooperon}
\end{center}
\end{figure}

\noindent
Remarkably, this mass range substantially corresponds to the $Z$-pole region which, as one can see in fig.(\ref{Fig:sigmasd}), is not yet excluded by LUX and 
respect $90$ \% annihilation into $b \bar b$ final state, provided that the interactions of the DM with the $Z$ are dominated by its axial component. We thus 
show in  fig.(\ref{Fig:hooperon}) the region in the parameter space ($V_\chi$, $A_\chi$) satisfying the correct DM relic density and at the same time accounting 
for the GC excess, according the determination of \cite{Calore:2014xka}.    

\noindent
We clearly see that it is not possible to satisfy both requirements, while being in agreement with LUX data. This is due to the fact that direct detection constraints can be satisfied only for $|A_\chi| \gg |V_\chi|$. In this regime, as discussed above, the present time DM annihilation cross-section is very different from the one at cosmological times. Their ratio can be estimated, using eq.(\ref{Eq:sigvff}), as\footnote{Notice that expression~(\ref{eq:GCth_ratio}) is not strictly valid at resonance, i.e. $m_\chi \sim \frac{m_Z}{2}$. In this case the ratio can be only computed numerically and his value is even lower than the one determined analytically.}:
\begin{align}
\label{eq:GCth_ratio}
& \frac{\langle \sigma v \rangle_{v \rightarrow 0}}{\langle \sigma v \rangle_{\rm f.o.}}\approx \frac{3}{2 v_{\rm f.o.}^2}\frac{m_b^2}{m_\chi^2}\frac{{\left(m_Z^2-4 m_\chi^2\right)}^2}{m_Z^4}\nonumber\\
& \frac{|A_b|^2}{\sum_{m_\chi > m_f} \left(|V_f|^2+|A_f|^2\right)}
\simeq O\left(10^{-3}\right)
\end{align} 
The value of this ratio is too low to be compatible with the Galactic Center signal. Indeed in order to reproduce it a value of the present time annihilation cross-section similar to the cosmologically one is required, i.e. $\frac{\langle \sigma v \rangle_{v \rightarrow 0}}{\langle \sigma v \rangle_{\rm f.o.}} \simeq 1$.

\subsection{Accelerator constraints}

\noindent
The Z-portal scenario could be in principle tested at LHC through searches of associated production events, like monojet events. However the expected prospects are rather poor. Indeed the production cross-section are largely suppressed by the current missing energy cuts and by kinematics at high DM masses \cite{Fox:2012ru,Buchmueller:2014yoa,Khachatryan:2014rra}. Moreover it would be problematic to disentangle the signal for background originated by $Z \rightarrow \bar\nu \nu$. This expectation is confirmed by analysis like the one presented in \cite{deSimone:2014pda,Buchmueller:2014yoa}, which show much less stringent bounds on $A_\chi$ with respect to dark matter observables. 

\noindent
Z-portal models in the low DM mass region could be probed instead at future $e^+ e^-$ collider, in which an improvement in the measure of the $Z$ invisible width is feasible \cite{Richard:2014vfa}.

\subsection{Examples of models}

\noindent
As already mentioned in the first section, the lagrangian written in eq.(\ref{eq:Zlagrangian}), should be regarded as just the low energy limit of 
a UV-complete model. A covariant formulation can be performed in terms of effective operators. A very simple choice, although not unique~(see e.g. \cite{Nagata:2014aoa} for an alterative choice in the case in which the DM is assumed to be charged under $SU(2) \times U(1)$), is:
\begin{equation}
\mathcal{L}=\frac{i g}{\Lambda^2} H^{\dagger}D_\mu H \left[\overline{\chi}\gamma^\mu \left(v_\chi+a_\chi \gamma_5\right)\chi\right]
\end{equation}
where $H$ is the Standard Model Higgs doublet.   
After EW symmetry breaking we have $H^{\dagger}D_\mu H \rightarrow \frac{v_h^2}{4 \cos\theta_W} Z_\mu$ and we thus obtain~(\ref{eq:Zlagrangian}) 
by defining $V_{\chi}=\frac{v_h^2}{\Lambda^2}\ v_\chi$ and $A_\chi$ analogously. Notice this operator also adds new contribution, from a 
four-field contact interaction, to the amplitude of the $\overline{\chi}\chi \rightarrow Zh$ process as well as a new annihilation channel into 
two Higgs bosons with cross-section:
\begin{equation}
\langle \sigma v \rangle_{\rm hh}\approx\frac{g^2}{4 \pi \cos\theta^2_W}|V_\chi|^2 \frac{m_\chi^2}{\Lambda^2}
\end{equation}
Alternatively a renormalizable interaction between the DM and the Z boson can be obtained in model with an extra $U(1)$ gauge symmetry. The DM originally 
interacts with the gauge bosons of this extra symmetry in the conventional form $\overline{\chi}\gamma^\mu D_{\mu} \chi$. At the same time the kinetic 
mixing term $\delta B_{\mu \nu}B^{'\,\mu \nu}$ is in general allowed by gauge invariance. After EW symmetry breaking an interaction between the $Z$-boson and the DM is induced. 
This would be the only relevant interaction of the dark matter with the visible sector in the case $m_{Z^{'}} \gg m_Z$. The lagrangian~(\ref{eq:Zlagrangian}) would mimic the kinetic mixing scenario provided 
that $V_\chi=4 q_D \frac{g_D}{g} \tan\theta_W \delta \frac{m_Z}{m_{Z^{'}}}$ and $A_\chi=0$ where $g_D$ and $q_D$ are, respectively, the new gauge 
coupling and the charge of the DM under the new gauge group. As evident in this last case we would have a rather different phenomenology with 
respect to the one discussed in this paper and would not be possible to a viable DM scenario excepted for DM masses in the multi-TeV range. Notice that, in presence of the only vectorial coupling, the enhanchment factor $m_\chi^2/m_V^2$ would be present only in the $WW$ annihilation channel, however there are 
no problems with respect to unitarity because of the suppressed coupling.
Interactions between the $Z$-bosons and fermionic DM are also present in supersymmetric theories. Here the scenario of pure axial couplings is naturally realized since the conventional LSP, i.e. the neutralino, is a Majorana fermion. We remark in particular that recent studies \cite{Cheung:2014lqa,Cao:2014efa} have found that a partial contribution to the relic density from $Z$ mediated annihilation can accomodate a viable DM scenario accounting from the GC excess, although the latters require the presence of a light CP-odd higgs as mediator of the present time annihilations.

\section{Conclusion}

\noindent
In this work, we have studied the so--called $Z-$portal, where the mediator between the dark sector and the Standard Model
is the $Z-$boson. Our results are well summarized in the fig.(\ref{Fig:all}) and (\ref{Fig:sigmasd}).
We showed that in the case of a hierarchy between the axial and the vectorial part, a part of the parameter space
is still allowed. For $m_\chi \gtrsim 200$ GeV, a ratio $|V_\chi|/|A_\chi|\simeq 10^{-2}-10^{-1}$ is imposed by the combined constraints
of LUX, PLANCK, and FERMI collaborations, implying an almost purely axially coupled dark matter to the $Z$ boson. 
Within these constraints it is not possible to fit the galactic center gamma-ray signal since the required mass 
corresponds to a suppressed annihilation cross-section at present times.
For heavier masses of the dark matter candidate ($m_\chi \gtrsim $ 1 TeV), universal coupling 
($|A_\chi| \simeq |V_\chi|$) are still allowed. The next generation of experiments will be able to test/exclude completely the $Z-$portal extension for dark matter masses below 2-3 TeV.

\noindent {\bf Acknowledgements. }  The authors would like  to thank  A. Pukhov, E. Bragina, D. Hooper, G. Belanger and M. Goodsell for very fruitful discussions. The authors would also like to thank Emilian Dudas and Lucien Heurthier for 
inspiring and useful comments.
This  work was supported by the Spanish MICINN's
Consolider-Ingenio 2010 Programme  under grant  Multi-Dark {\bf CSD2009-00064}, and the contract {\bf FPA2010-17747}.
Y.M. is grateful to the Mainz Institute for Theoretical Physics
(MITP) for its hospitality and its partial support during the completion of this
work. Y.M.  and G.A. acknowledge partial support from the European Union FP7 ITN INVISIBLES (Marie
Curie Actions, PITN- GA-2011- 289442) and  the ERC advanced grants  
 MassTeV and  Higgs@LHC.

\onecolumngrid
\appendix

\section{DM-nucleon scattering cross-sections}

\noindent
The SI and SD components of the DM scattering cross-section with proton and neutrons are given by:
\begin{align}
& \sigma^p_{SI}=\frac{g^4 |V_\chi|^2 \mu_{\chi p}^2}{4 \cos^2\theta_W \pi m_{Z}^4}\frac{\sum_A \eta_A A
\bigg[ V_u \left(1 + \frac{Z}{A}\right) + V_d \left(2 - \frac{Z}{A}\right)  \biggr]^2}{\sum_A \eta_A A^2},\,\,\,\,\,\,\,\sigma^n_{SI}=\sigma^p_{SI}\frac{\mu_{\chi n}^2}{\mu_{\chi p}^2} \frac{\left(2 V_u+V_d\right)^2}{\left(V_u+2 V_d\right)^2} \nonumber\\
& \sigma^p_{SD}= \frac{3 g^4 \mu_{\chi p}^2 |A_\chi|^2}{4 \cos^2\theta_W \pi m_{Z}^4} \frac{\sum_A\eta_A
\biggl[ A_u(\Delta_u^p S_p^A 
+ \Delta_d^pS_n^A) + A_d \left( (\Delta_d^p + \Delta_s^p) S_p^A+ (\Delta_u^p + \Delta_s^p)S_n^A  \right) \biggr]^2}{\sum_A \eta_A (S_p^A + S_n^A)^2}, \nonumber\\
& \sigma^n_{SD}=\sigma^p_{SI}\frac{\mu_{\chi n}^2}{\mu_{\chi p}^2}\frac{\left(A_u \Delta^p_u+A_d \left(\Delta^p_d+\Delta^p_s\right)\right)^2}{\left(A_u \Delta^n_u+A_d \left(\Delta^n_d+\Delta^n_s\right)\right)^2}
\end{align}
where the extra factors depending on $Z$ and $A$, for the SI case, and $S_{p,n}^A$, i.e. the proton and neutron contributions to the nucleus spin, in the SD case, have been introduced in order to directly compare these cross-sections with the experimental results, since these customarily assume equal interactions of the DM with proton and neutrons. This is not the case of Z-portal scenarios. The sums run over the isotopes, with relative abundace $\eta_A$, of the target material \cite{Feng:2013vod}. In the case of SD cross-section only the isotopes with odd number of nucleons give non-zero contribution.

\subsection{Annihilation Cross sections}

\noindent
The DM annihilates into fermion pairs. We identify the following three contributions to the pair annihilation cross-section:

\begin{align}
& {\sigma (s)}_{\rm VV}=\frac{1}{192 \pi \cos^2\theta_W } n_c^f |V_\chi|^2 |V_f|^2\frac{g^4 s}{{\left(s-m_{Z}^2\right)}^2+m_{Z}^2 \Gamma_{Z}^2}\frac{\sqrt{1-4 \mu_f^2}}{\sqrt{1-4 \mu_\chi^2}} \left[\left(1+2 \mu_\chi^2\right)\left(1+2 \mu_f^2\right)\right] \nonumber\\
& {\sigma(s)}_{\rm VA}=\frac{1}{192 \pi \cos^2\theta_W} n_c^f \frac{g^4 s}{{\left(s-m_{Z}^2\right)}^2+m_{Z}^2 \Gamma_{Z}^2}\frac{\sqrt{1-4 \mu_f^2}}{\sqrt{1-4\mu^2_\chi}}\left[|V_\chi|^2 |A_f|^2\left(1-4 \mu_f^2\right)\left(1+2 \mu_\chi^2\right)+|A_\chi|^2 |V_f|^2 \beta^2 \left(1+2\mu_f^2\right)\right] \nonumber\\
&  {\sigma (s)}_{\rm AA}=\frac{1}{192 \pi \cos^2\theta_W} n_c^f \frac{g^4 s}{{\left(s-m_{Z}^2\right)}^2+m_{Z}^2 \Gamma_{Z}^2}\frac{\sqrt{1-4 \mu_f^2}}{\sqrt{1-4\mu_\chi^2}}|A_\chi|^2 |A_f|^2\left[\beta^2+28 \mu_f^2 \mu_\chi^2+12 \mu_f^2 \mu_\chi^2 \frac{s^2}{m_{Z}^4}-4 \mu_f^2 \left(1+6 \mu_\chi^2 \frac{s}{m_{Z}}\right)\right]
\label{Eq:sigvff1}
\end{align} 
where $\mu_{f,\chi}=\frac{m_{f,\chi}}{\sqrt{s}}$ and $\beta=\sqrt{1-4 \mu_{\chi}^2}$.

\noindent
We can correlate these contributions to the annihilation cross-section to the scattering cross section of the DM as:

\begin{align}     
& {\sigma (s)}_{\rm VV}=\sum_{m_f < m_\chi} n_c^f \frac{1}{48\mu^2_{\chi-p}} |V_f|^2\frac{\sigma_{\chi-p}^{\rm SI}}{\alpha^{\rm SI}_{Z,A}}\frac{m_{Z}^4 s}{{\left(s-m_{Z}^2\right)}^2+m_{Z}^2 \Gamma_{Z}^2}\frac{\sqrt{1-4 \mu_f^2}}{\sqrt{1-4\mu_\chi^2}}\left[\left(1+2 \mu_\chi^2\right)\left(1+2 \mu_f^2\right)\right] \nonumber\\
& {\sigma (s)}_{\rm VA}=\sum_{m_f < m_\chi} n_c^f \frac{1}{48\mu^2_{\chi-p}} \frac{m_{Z}^4 s}{{\left(s-m_{Z}^2\right)}^2+m_{Z}^2 \Gamma_{Z}^2}\frac{\sqrt{1-4 \mu_f^2}}{\sqrt{1-4\mu_\chi^2}}\nonumber\\
&\times \left[\frac{\sigma_{\chi-p}^{\rm SI}}{\alpha^{\rm SI}_{Z,A}} |A_f|^2\left(1-4 \mu_f^2\right)\left(1+2 \mu_\chi^2\right)+\frac{\sigma_{\chi-p}^{\rm SD}}{3\alpha^{\rm SD}_{Z,A}} |V_f|^2 \beta^2 \left(1+2\mu_f^2\right)\right] \nonumber\\
&  {\sigma (s)}_{\rm AA}=\sum_{m_f < m_\chi} n_c^f\frac{1}{48\mu^2_{\chi-p}} \frac{m_{Z}^4 s}{{\left(s-m_{Z}^2\right)}^2+m_{Z}^2 \Gamma_{Z}^2}\frac{\sqrt{1-4 \mu_f^2}}{\sqrt{1-4\mu_\chi^2}}\nonumber\\
&\times \frac{\sigma_{\chi-p}^{\rm SD}}{3\alpha^{\rm SD}_{Z,A}} |A_f|^2\left[\beta^2+28 \mu_f^2 \mu_\chi^2+12 \mu_f^2 \mu_\chi^2 \frac{s^2}{m_{Z}^4}-4 \mu_f^2 \left(1+6\mu_\chi^2 \frac{s}{m_{Z}}\right)\right]
\end{align}

\noindent
The thermal average is defined as:
\begin{equation}
\langle \sigma v \rangle=\frac{1}{8 m_\chi^4 T {K_2\left(\frac{m_\chi}{T}\right)}^2}\int_{4 m_\chi^2}^{\infty} ds \sigma(s) \sqrt{s} \left(s-4 m_\chi^2\right)K_1\left(\frac{\sqrt{s}}{T}\right)
\end{equation}

\subsubsection{Velocity expansion of the annihilation cross section}

\noindent
Away from resonances, a manageable analytical expression for the thermally averaged pair annihilation cross section is obtained by performing the formal velocity expansion, in the non-relativistic limit, as defined in \cite{Gondolo:1990dk}. The thermally averaged cross section can be computed as:
\begin{equation}
\label{eq:Gondolo_int}
\langle \sigma v \rangle = \frac{2 x^{3/2}}{\pi^{1/2}}\int_0^{\infty} \sigma v_{\rm lab}\epsilon^{1/2}e^{-\epsilon x}
\end{equation}
where:
\begin{align}
& v_{\rm lab}=\frac{2 \epsilon^{1/2}{\left(1+\epsilon\right)}^{1/2}}{\left(1+2\epsilon\right)} \nonumber\\
& \epsilon=\frac{s-4 m_\chi^2}{4 m_\chi^2}\,\,\,\,\,\,x=\frac{m_\chi}{T}
\end{align}
This kind of integral can be analytically computed by considering an expansion in series of $\epsilon$ of $\sigma v_{\rm lab}$, namely:
\begin{equation}
\sigma v_{\rm lab}=a_0 +a_1 \epsilon +a_2 \epsilon^2 \cdots
\end{equation}
for our computation the first two terms of the expansion are relevant. These are given by:

\begin{align}
& a_0=\frac{g^4}{16 \cos^2 \theta_W}\frac{2 \sqrt{m_\chi^2-m_f^2} \left(|A_f|^2 |A_\chi|^2 m_f^2 \left(m_{Z}^2-4 m_\chi^2\right)^2+m_{Z}^4 |V_\chi|^2 \left(2
   |A_f|^2 \left(m_\chi^2-m_f^2\right)+|V_f|^2 \left(m_f^2+2 m_\chi^2\right)\right)\right)}{4 \pi  m_{Z}^4 m_\chi \left(m_{Z}^2-4
   m_\chi^2\right)^2} \nonumber\\
  \vspace{0.5 cm} 
& a_1=-\frac{g^4}{16 \cos^2 \theta_W}\frac{1}{12 \pi  m_{Z}^4 m_\chi \sqrt{m_\chi
  ^2-m_f^2} \left(m_{Z}^2-4 m_\chi^2\right)^3}   
  \left(|A_f|^2 \left(2 m_{Z}^4 |V_\chi|^2 (m_f-m_\chi) (m_f+m_\chi) \left(-2 m_\chi^2 \left(46
   m_f^2+m_{Z}^2\right) \right.\right. \right. \nonumber\\
   & \left. \left. \left. +11 m_f^2 m_{Z}^2+56 m_\chi^4\right)-|A_\chi|^2 \left(m_{Z}^2-4 m_\chi^2\right) \left(23 m_f^4
   m_{Z}^4-4 m_f^2 m_{Z}^2 m_\chi^2 \left(30 m_f^2+7 m_{Z}^2\right) \right. \right. \right.\nonumber\\
   & \left. \left. \left. -192 m_f^2 m_\chi^6+8 m_\chi^4 \left(30 m_f^4+12
   m_f^2 m_{Z}^2+m_{Z}^4\right)\right)\right)+m_{Z}^4 |V_f|^2 \left(4 |A_\chi|^2 \left(m_f^4+m_f^2 m_\chi^2-2 m_\chi^4\right) 
   \left(m_{Z}^2-4 m_\chi^2\right)\right. \right. \nonumber\\
   & \left. \left.+|V_\chi|^2 \left(-11 m_f^4 m_{Z}^2+4 m_\chi^4 \left(14 m_f^2+m_{Z}^2\right)-2
   m_f^2 m_\chi^2 \left(m_{Z}^2-46 m_f^2\right)-112 m_\chi^6\right)\right)\right)
   \end{align}

\noindent   
As evident, the contribution to the $0th-$order term proportional to the DM axial coupling is proportional to the SM fermion mass, and thus suppressed, while the corresponding $1st$-order term not. A reliable computation thus requires an expansion featuring at least the first two orders. 
Plugging the result above into the integral~(\ref{eq:Gondolo_int}) we obtain the thermally averaged annihilation cross section up to the $O(v^2)$:

\begin{align}
& \langle \sigma v \rangle_{\rm ff}=\frac{g^4}{16 \cos^2\theta_W} \sum_{m_f < m_\chi} N_c^f\sqrt{m_\chi^2-m_f^2}\nonumber\\
&\frac{2  \left(|A_f|^2 |A_\chi|^2 m_f^2 \left(m_{Z}^2-4 m_\chi^2\right)^2+m_{Z}^4 |V_\chi|^2 \left(2 |A_f|^2 \left(m_\chi^2-m_f^2\right)+|V_f|^2 \left(m_f^2+2 m_\chi^2\right)\right)\right)}{4 \pi m_\chi m_{Z}^4 \left(m_{Z}^2-4 m_\chi^2\right)^2}\nonumber\\
&-\frac{1}{24 \pi  m_\chi m_{Z}^4 \sqrt{m_\chi^2-m_f^2} \left(m_{Z}^2-4 m_\chi^2\right)^3}v^2 \left(|A_f|^2 \left(2
   m_{Z}^4 |V_\chi|^2 (m_f-m_\chi) (m_f+m_\chi)\right. \right.\nonumber\\
& \left. \left.\left(-2 m_\chi^2 \left(46 m_f^2+m_{Z}^2\right)+11 m_f^2 m_{Z}^2+56 m_\chi^4\right)-|A_\chi|^2 \left(m_{Z}^2-4 m_\chi^2\right)\right.\right.\nonumber\\
& \left.\left. \left(23 m_f^4 m_{Z}^4-192 m_f^2 m_\chi^6-4 m_f^2 m_\chi^2 m_{Z}^2 \left(30 m_f^2+7
   m_{Z}^2\right)+8 m_\chi^4 \left(30 m_f^4+12 m_f^2 m_{Z}^2+m_{Z}^4\right)\right)\right)\right.\nonumber\\
&\left.+m_{Z}^4 |V_f|^2 \left(4 |A_\chi|^2 \left(m_f^4+m_f^2 m_\chi^2-2 m_\chi^4\right) \left(m_{Z}^2-4 m_\chi^2\right)\right.\right.\nonumber\\
&\left.\left.+|V_\chi|^2 \left(-11 m_f^4 m_{Z}^2+4 m_\chi^4 \left(14 m_f^2+m_{Z}^2\right)-2 m_f^2 m_\chi^2 \left(m_{Z}^2-46 m_f^2\right)-112 m_\chi^6\right)\right)\right)
\label{Eq:sigvff2}
\end{align}	

\noindent	
where $v=\sqrt{3 T/m_\chi}$.

\noindent
The other annihilation channels can be evaluated through an analogous procedure. By increasing the DM mass first opens the annihilation channel into two W-bosons whose thermally averaged cross-section is:

\begin{align}
& \langle \sigma v \rangle_{W^{+}W^{-}}=\frac{\pi \alpha_{\rm e.m.}}{4 \tan\theta_W \cos^2\theta_W}g^2\tilde{\sigma}_{W^{+}W^{-}} \nonumber\\
&\tilde{\sigma}_{W^{+}W^{-}}= \frac{|V_\chi|^2 
   \sqrt{m_\chi^2-m_W^2} \left(-3 m_W^6-17
   m_W^4 m_\chi^2+16 m_W^2 m_\chi^4+4
   m_\chi^6\right)}{4 \pi  m_W^4 m_\chi
   \left(m_Z^2-4 m_\chi^2\right)^2} \nonumber\\
& \frac{v^2 \sqrt{m_\chi^2-m_{W}^2}}{48 \pi  m_W^4 m_\chi
   \left(m_Z^2-4 m_\chi^2\right)^3}
    \left(4 |A_\chi|^2
   \left(-3 m_W^6-17 m_W^4 m_\chi^2+16 m_W^2
   m_\chi^4+4 m_\chi^6\right) \left(m_Z^2-4
   m_\chi^2\right)\right.\nonumber\\
 &\left.+   |V_\chi|^2 \left(33 m_W^6
   m_Z^2+8 m_\chi^6 \left(58 m_W^2+5
   m_Z^2\right)+4 m_W^2 m_\chi^4 \left(19
   m_Z^2-298 m_W^2\right)+2 m_W^4 m_\chi^2
   \left(47 m_Z^2-138 m_W^2\right)+32 m_\chi^8\right)\right) 
   \end{align}

\noindent
In the case $m_{\chi} > m_{Z}$ we have to add the contribution of the $\chi \chi \rightarrow ZZ$ annihilation:

\begin{align}
&\langle \sigma v \rangle_{ZZ}=
\frac{g^2}{16 \cos^2 \theta_W} \frac{\left(m_\chi^2-m_{Z}^2\right)^{3/2} \left(|A_\chi|^4 m_{Z}^2+2 |A_\chi|^2 |V_\chi|^2 \left(4 m_\chi^2-3
   m_{Z}^2\right)+m_{Z}^2 |V_\chi|^4\right)}{2 \pi  m_\chi \left(m_{Z}^3-2 m_\chi^2 m_{Z}\right)^2}+\nonumber\\
&\frac{1}{24 \pi 
   m_\chi \left(m_{Z}^3-2 m_\chi^2 m_{Z}\right)^4}v^2
   \sqrt{m_\chi^2-m_{Z}^2} \left(|A_\chi|^4 \left(128 m_\chi^{10}+23 m_{Z}^{10}-118 m_\chi^2 m_{Z}^8+172
   m_\chi^4 m_{Z}^6\right.\right.\nonumber\\
& \left.\left.+32 m_\chi^6 m_{Z}^4-192 m_\chi^8 m_{Z}^2\right)\right.\nonumber\\
&\left. -2 |A_\chi|^2 m_{Z}^2 |V_\chi|^2
   \left(160 m_\chi^8+21 m_{Z}^8-182 m_\chi^2 m_{Z}^6+508 m_\chi^4 m_{Z}^4-528 m_\chi^6
   m_{Z}^2\right)\right. \nonumber\\
&\left. +m_{Z}^6 |V_\chi|^4 \left(76 m_\chi^4+23 m_{Z}^4-66 m_\chi^2 m_{Z}^2\right)\right)
\end{align}	

\noindent
For the highest values of the DM masses we have finally to add the contribution of $Zh$ channel. The corresponding cross-section is given by:

\begin{align}
& \langle \sigma v \rangle_{\rm Zh}=\frac{4 m_Z^4}{v_h^2}\frac{g^2}{16 \cos^2\theta_W}\tilde{\sigma}_{\rm Zh} \nonumber\\
& \tilde{\sigma}_{\rm Zh}=\frac{\sqrt{m_h^4-2 m_h^2 \left(4 m_\chi^2+m_Z^2\right)+\left(m_Z^2-4 m_\chi^2\right)^2}}{3072 \pi  m_\chi^4 m_Z^6} 
\left(3 |A_\chi|^2 \left(m_h^4-2 m_h^2 \left(4 m_\chi^2
+m_Z^2\right)+\left(m_Z^2-4 m_\chi^2\right)^2\right)\right.\nonumber\\
& \left. +\frac{3 m_Z^4 |V_\chi|^2 \left(-8 m_\chi^2 \left(m_h^2-5
   m_Z^2\right)+\left(m_h^2-m_Z^2\right)^2+16 m_\chi^4\right)}{\left(m_Z^2-4 m_\chi^2\right)^2}\right)\nonumber\\
& -\frac{v^2 \sqrt{m_h^4-2 m_h^2 \left(4 m_\chi^2+m_Z^2\right)+\left(m_Z^2-4 m_\chi^2\right)^2}}{3072 \pi  m_\chi^4 m_Z^6 \left(m_Z^2-4 m_\chi^2\right)^3
   \left((m_h-m_Z)^2-4 m_\chi^2\right) \left((m_h+m_Z)^2-4 m_\chi^2\right)} \nonumber\\
&	\left(|A_\chi|^2 \left(m_Z^2-4 m_\chi^2\right) \left(-96 m_\chi^6 \left(5
   m_h^2+7 m_Z^2\right)+5 m_Z^4 \left(m_h^2-m_Z^2\right)^2+8 m_\chi^4 \left(12 m_h^4+6 m_h^2 m_Z^2+43 m_Z^4\right)\right.\right.\nonumber\\
& \left.\left. -2 m_\chi^2 m_Z^2 \left(24
   m_h^4-37 m_h^2 m_Z^2+59 m_Z^4\right)+384 m_\chi^8\right) \left(m_h^4-2 m_h^2 \left(4 m_\chi^2+m_Z^2\right)+\left(m_Z^2-4 m_\chi^2\right)^2\right)\right.\nonumber\\
& +\left. m_Z^4 |V_\chi |^2 \left(128 m_\chi^8 \left(37 m_h^2-82 m_Z^2\right)+5 m_Z^2 \left(m_h^2-m_Z^2\right)^4+32 m_\chi^6 \left(-69
   m_h^4+217 m_h^2 m_Z^2+242 m_Z^4\right)\right.\right.\nonumber\\
& \left.\left. +2 m_\chi^2 \left(m_h^2-m_Z^2\right)^2 \left(-16 m_h^4+m_h^2 m_Z^2+37 m_Z^4\right)+8 m_\chi^4 \left(55
   m_h^6-178 m_h^4 m_Z^2+147 m_h^2 m_Z^4-200 m_Z^6\right)\right.\right.\nonumber\\
& \left.\left.   -3584 m_\chi^{10}\right)\right)
\end{align}	

\noindent
and $v_h$ is the {\it vev} of the Higgs.

\vspace{1cm}

\twocolumngrid

\end{document}